\documentclass[a4paper,10pt]{article}
\usepackage[cp1251]{inputenc}
\usepackage[english]{babel}
\usepackage{amsmath}
\usepackage{amssymb}
\usepackage{graphicx}
\usepackage[square,numbers,comma,sort&compress]{natbib}

\begin{document}
%\vspace{-0.25cm}
\twocolumn[\fontsize{6pt}{7pt}\selectfont\textsl{ISSN 0021-3640, JETP Letters, 2020, Vol. 111, No. 2, pp. 90--95. \textcopyright\,Pleiades Publishing, Inc., 2020.\\ Original Russian Text \textcopyright\,The Author(s), 2020, published in Pis'ma v Zhurnal Eksperimental'noi i Teoreticheskoi Fiziki, 2020, Vol. 111, No. 2, pp. 80--85.}

\vspace{0.18cm}

\begin{center}
\rule{7cm}{0.4pt}\hspace{3cm}\rule{7cm}{0.4pt}

\vspace{-0.15cm}
\rule{7cm}{0.4pt}\hspace{3cm}\rule{7cm}{0.4pt}

\normalsize
\vspace{-0.55cm}
{\bf CONDENSED\\
MATTER}

\vspace{0.5cm}

\LARGE{\bf Electron--Hole Liquid in Monolayer Transition Metal\\
Dichalcogenide Heterostructures}

\vspace{0.5cm}

\large{\bf P.\,L. Pekh$^{a,\,*}$, P.\,V. Ratnikov$^b$, and A.\,P. Silin$^{a,\,c}$}

\vspace{0.15cm}

\normalsize

\textit{$^a$\,Lebedev Physical Institute, Russian Academy of Sciences, Moscow, 119991 Russia\\
$^b$\,Prokhorov General Physics Institute, Russian Academy of Sciences, ul. Vavilova 38, Moscow, 117942 Russia\\
$^c$\,Moscow Institute of Physics and Technology (National Research University), Dolgoprudnyi, Moscow region, 141700 Russia\\
$^*$e-mail: pavel.pekh@phystech.edu}

\vspace{0.25cm}

Received October 1, 2019; revised November 19, 2019; accepted December 5, 2019
\end{center}

\vspace{0.1cm}
\begin{list}{}
{\rightmargin=0.85cm\leftmargin=0.85cm}
\item
\small{Monolayer films of transition metal dichalcogenides (in particular, MoS2, MoSe2, WS2, and WSe2) can be considered as ideal systems for the studies of high-temperature electron--hole liquids. The quasi-two-dimensional nature of electrons and holes ensures their stronger interaction as compared to that in bulk semiconductors. The screening of the Coulomb interaction in monolayer heterostructures is significantly reduced, since it is determined by the permittivities of the environment (e.g., vacuum and substrate), which are much lower than those characteristic of the films of transition metal dichalcogenides. The multivalley structure of the energy spectrum of charge carriers in transition metal dichalcogenides significantly reduces the kinetic energy, resulting in the increase in the equilibrium density and binding energy of the electron--hole liquid. The binding energy of the electron--hole liquid and its equilibrium density are determined. It is shown that the two-dimensional Coulomb potential should be used in the calculations for the electron--hole liquid.}

\vspace{0.5cm}

\normalsize{\bf DOI}: 10.1134/S0021364020020101
\end{list}\vspace{0.5cm}]

\begin{center}
1. INTRODUCTION
\end{center}

An increased interest in the studies of graphene as a promising material for nanoelectronics \cite{RS2018} has led to the development of novel two-dimensional (2D)
materials, such as monolayers of hexagonal boron nitride, black phosphorus, and transition metal dichalcogenides (TMDs) \cite{Miro2014}. Recently, vertical (van
der Waals) heterostructures, where various 2D materials are combined in a given sequence, have been actively studied \cite{Geim2013}.

Monomolecular layers of TMDs with the chemical formula $MX_2$, where $M$ is a transition metal and $X$ is a chalcogen, are of a particular interest. Among them,
semiconductors with Group VI metal atoms (M = Mo, W) and S, Se, and Te chalcogens are the most studied. Bulk layered TMDs (e.g., MoS$_2$, WS$_2$, MoSe$_2$, and WSe$_2$) have an indirect energy gap $E_g\sim1$ eV \cite{Bulaevskii1975, Bulaevskii1976}, whereas the corresponding monomolecular layers are direct-gap semiconductors with $E_g$ about 2 eV \cite{Durnev2018}.

Many bulk TMD samples were obtained as early as the 1960s \cite{Wilson1969}. Even at that time, their electron properties were already actively investigated \cite{Kalihman1972, Silin1978}. In particular, some TMDs (M = Nb, Ta, Ti, Mo; X = S, Se) exhibit a low-temperature transition to a superconducting state. The structure, synthesis, properties, and applications of TMDs are described in detail in the recent review \cite{Chernozatonskii2018}.

The optical properties of monomolecular TMD layers are mainly determined by excitons and trions. The exciton binding energy Ex in TMDs is of the order
of hundreds of meV (e.g., in MoS$_2$ monolayers, $E_x=420$ meV \cite{Yu2016}), and the corresponding energy for trions is equal to tens of meV \cite{Durnev2018}.

These facts suggest that the arrays involving TMD monolayers are ideal systems for the studies of a hightemperature electron--hole liquid (EHL). The energy
of one electron-hole pair in the EHL is $\left|E_\text{EHL}\right|\sim E_x$ and the critical temperature of the gas--liquid phase transition is $T_c\sim0.1\left|E_\text{EHL}\right|$ \cite{Andryushin1977a, Rice1977, JeffrisKeldysh1988, Tikhodeev1985, Sibeldin2016, Sibeldin2017}; therefore, it can be
expected that the EHL will be observed in TMD monolayers even at room temperature. In MoS$_2$ monolayers, a high-temperature strongly coupled
EHL with $T_c\simeq500$ K has been already observed \cite{Yu2019}.

In this work, we investigate the possibility of forming the EHL in monolayers of multivalley semiconductors \cite{Silin1978, Andryushin1979b}. We consider a thin film of a model multivalley semiconductor deposited in vacuum on a dielectric substrate. We assume that the semiconductor has a fairly wide band gap $E_g\gg\left|E_\text{EHL}\right|$ and use the single-band approximation. The semiconductor has a large equal number of equivalent electron $\nu_e$ and hole $\nu_h$ valleys, $\nu_e=\nu_h=\nu\gg1$ with the effective masses $m_e$ and $m_h$ of the electron and hole, respectively. The multivalley structure can be due to the presence of several molecular monolayers in the film.

In \cite{Andryushin1976a}, it was shown that the interaction energy of charge carriers belonging to different valleys dominates in such a system at $\nu\gg1$. The equilibrium density $n_\text{EHL}$ of the EHL and the energy $E_\text{EHL}$ corresponding to this density increase drastically. This increase in
the density justifies the use of the random phase approximation to calculate the correlation energy.

\begin{center}
2. MODEL
\end{center}

We study the 2D electron--hole system described by the Hamiltonian \cite{Andryushin1976b, Andryushin1976c}
\begin{equation*}
\begin{split}
&\widehat{H}=\sum_{\mathbf{p}sk}^{\nu_e}\varepsilon^e_{sk}(\mathbf{p})a^\dag_{\mathbf{p}sk} a_{\mathbf{p}sk}+\sum_{\mathbf{p}sl}^{\nu_h}\varepsilon^h_{sl}(\mathbf{p})b^\dag_{\mathbf{p}sl}b_{\mathbf{p}sl}+\\
&+\frac{1}{2}\sum_{\mathbf{p}\mathbf{p}^\prime\mathbf{q}ss^\prime}V(\mathbf{q})\left\{\sum_{kk^\prime}^{\nu_e}a^\dag_{\mathbf{p}sk}a^\dag_{\mathbf{p}^\prime s^\prime k^\prime}a_{\mathbf{p}^\prime+\mathbf{q}s^\prime k^\prime}a_{\mathbf{p}-\mathbf{q}sk}+\right.\\
\end{split}
\end{equation*}
\begin{equation}\label{Ham}
+\sum_{ll^\prime}^{\nu_h}b^\dag_{\mathbf{p}sl}b^\dag_{\mathbf{p}^\prime s^\prime l^\prime}b_{\mathbf{p}^\prime+\mathbf{q}s^\prime l^\prime}b_{\mathbf{p}-\mathbf{q}sl}-
\end{equation}
\begin{equation*}
\left.-2\sum_{kl}^{\nu_e\nu_h}a^\dag_{\mathbf{p}sk}b^\dag_{\mathbf{p}^\prime s^\prime l}b_{\mathbf{p}^\prime+\mathbf{q}s^\prime l}a_{\mathbf{p}-\mathbf{q}sk}\right\}.
\end{equation*}
Here, $a_{\mathbf{p}sk}$ ($a^\dag_{\mathbf{p}sk}$) and $b_{\mathbf{p}sl}$ ($b^\dag_{\mathbf{p}sl}$) are fermion annihilation (creation) operators for an electron and a hole having the crystal momentum $\mathbf{p}$ and spin projection $s$, which are located in $k$th and $l$th valleys, respectively. The dispersion relations of electrons and holes are
\begin{equation}\label{disp_lows}
\varepsilon^e_{sk}(\mathbf{p})=\frac{\mathbf{p}^2}{2(1+\sigma)m},~\varepsilon^h_{sl}(\mathbf{p})=\frac{\mathbf{p}^2}{2(1+1/\sigma)m},
\end{equation}
where $\sigma=m_e/m_h$ and $m=m_em_h/(m_e+m_h)$ is the reduced mass of the electron and hole. Usually, $\sigma\leq1$.

The Coulomb interaction in the films of a finite thickness is described by the Keldysh potential \cite{Rytova1967,Keldysh1979}
\begin{equation}\label{keldysh_int}
V(\mathbf{q})=\frac{2\pi\widetilde{e}^2}{|\mathbf{q}|(1+r_0|\mathbf{q}|)}.
\end{equation}
Here, $\widetilde{e}^2=e^2/\epsilon_\text{eff}$, where $\epsilon_\text{eff}=(\epsilon_1+\epsilon_2)/2$ is the effective permittivity of the media adjacent to the film (e.g., $\epsilon_1=1$ is the permittivity of vacuum and $\epsilon_2$ is the permittivity of the substrate), $r_0=d/2\delta$ is the screening length, $\delta=\epsilon_\text{eff}/\epsilon$ ($\epsilon$ is the permittivity of the substrate material), and $d$ is the film thickness.

For the monolayer film ($d\rightarrow0$), Eq. \eqref{keldysh_int} leads to the following conventional expression for the Coulomb interaction of charge carriers in the 2D system:
\begin{equation}\label{coloumb_2d}
V(\mathbf{q})=\frac{2\pi\widetilde{e}^2}{|\mathbf{q}|}.
\end{equation}

In terms of macroscopic electrodynamics, the screening of the Coulomb interaction of charge carriers is determined by the permittivities of media surrounding the film, since the electric field lines turn out to be outside the film. The introduction of the permittivity $\epsilon$ for monolayer TMD films, as well as for graphene \cite{RS2018}, has no physical meaning.

The nonzero term $r_0|\mathbf{q}|$ in the denominator of Eq. \eqref{keldysh_int} is introduced in order to explain the significant
deviation of the energies of several first exciton levels from the Rydberg series \cite{Durnev2018}.

First, we use potential \eqref{coloumb_2d} (the beginning of Section~3, Subsections 3.1 and 3.2). Then, to verify the validity of using potential \eqref{coloumb_2d}, as well as to compare our results to experimental data, we carry out calculations using the Keldysh potential (Subsection 3.3). The calculation results for the monolayer MoS2 film with both potentials are compared in Section 5.

\begin{center}
3. GROUND STATE ENERGY
\end{center}

The ground state energy per electron--hole pair in the 2D EHL can be written as \cite{Andryushin1976b, Andryushin1976c}
\begin{equation}\label{E_gs}
E_\text{gs}=E_\text{kin}+E_\text{exc}+E_\text{cor}.
\end{equation}
where
\begin{equation}\label{E_kin}
E_\text{kin}=\frac{\hbar^2\pi n_\text{2D}}{2m\nu}=\frac{1}{r_s^2}
\end{equation}
is the average kinetic energy, where $n_\text{2D}$ is the 2D electron and hole density;
\begin{equation}\label{Eexch}
E_\text{exc}=-\frac{8\sqrt{2}\widetilde{e}^2}{3\sqrt{\pi}}\sqrt{\frac{n_\text{2D}}{\nu}}=-\frac{8\sqrt{2}}{3\pi r_s}.
\end{equation}
is the exchange energy; and $E_\text{cor}$ is the correlation energy defined below. Here, $r_s=\sqrt{\nu/\pi n_\text{2D}}$ is the dimensionless interparticle distance. The Fermi wave vector is $q_F=\sqrt{2\pi n_\text{2D}/\nu}=\sqrt{2}/r_s$. Here and further on, we use the system of units in which the binding energy and radius of the 2D exciton are equal to unity: $E_x=2m\widetilde{e}^4/\hbar^2=1$ and $a_x=\hbar^2/2m\widetilde{e}^2=1$.

The main problem in determining the ground state energy of the EHL is the calculation of the correlation energy. In the simplest case of a single-valley semiconductor, it was calculated in \cite{Andryushin1976b, Andryushin1976c} by the Nosieres--Pines method. It was shown that, in contrast to the three-dimensional case, the 2D EHL appears to be more favorable in energy than the exciton gas even in the isotropic case. In this situation, the main contribution to $E_\text{cor}$ comes from the momentum transfer
exceeding the Fermi momentum.

The calculations of the EHL energy have been recently reported in \cite{Rustagi2018}. The correlation energy of the
electron gas in narrow-gap multivalley and layered semiconductors is calculated in \cite{Pechenik1996, Andryushin1996}. The Wannier-Mott excitons in heterostructures composed of narrow-gap semiconductors are addressed in \cite{Silin2000}.

In \cite{Babichenko2013}, the EHL in double quantum wells with spatially separated electrons and holes in multivalley semiconductors is studied. The energy and equilibrium density of the EHL are calculated at various distances between the electron and hole layers. The procedure for calculating the correlation energy of the 2D EHL at the spatial separation of electrons and holes is described in~\cite{Silin1983}.

\begin{center}
\textit{3.1. Calculation of the Correlation Energy\\
at a Finite Number of Valleys}
\end{center}

The correlation energy can be represented in the form of an integral over the momentum transfer \cite{Andryushin1976b, Andryushin1976c, Combescot1972, Andryushin1977b}
\begin{equation}\label{corr_e}
E_\text{corr}=\int\limits_0^\infty I(q)dq.
\end{equation}
At small (compared to $q_F$) q values, the function $I(q)$ is calculated in the random phase approximation, whereas at large q values, it is determined by the sum of diagrams in the second order of perturbation theory in terms of the interaction energy.

For any value of $\sigma$, the expansion of $I(q)$ at small $q$ is rather lengthy. We show here the result for the particular case of equal masses of electrons and holes ($\sigma=1$)
\begin{equation}\label{I_q}
I(q)=\begin{cases}
  -\frac{2\sqrt{2}}{\pi r_s}q+\frac{2^{1/4}}{r^{3/2}_s\nu^{1/2}}q^{3/2}-\frac{\pi+2}{2\pi r^2_s\nu}q^2+&\\
  +\frac{3}{2^{13/4}r^{5/2}_s\nu^{3/2}}q^{5/2}+\frac{r^2_s\nu^2-1}{6\pi\sqrt{2}r^3_s\nu^2}q^3, & \hspace{-0.5cm}q\ll1, \\
  -2(4\nu-1)/q^3, &  \hspace{-0.5cm}q\gg1.
\end{cases}
\end{equation}

In the intermediate range $q_1\leq q\leq q_2$, the function $I(q)$ is approximated by a segment of the tangent, as in \cite{Andryushin1976b, Andryushin1976c}. At $q\ll1$, we integrate expansion \eqref{I_q} from zero to the matching point $q_1\approx q_0$ ($q_0$ is the point corresponding to the minimum of the function $I(q)$), whereas the asymptotic expression at $q\gg1$ is integrated from the matching point $q_2$ to infinity. Adding the contribution to the integral from the intermediate range $\left(I(q_1)+I(q_2)\right)(q_2-q_1)/2$, we find
\begin{equation}\label{Ecorr}
\begin{split}
&E_\text{cor}=\left(-\frac{\sqrt{2}}{\pi r_s}q_0+\frac{1}{2^{3/4}r^{3/2}_s\nu^{1/2}}q^{3/2}_0-\frac{\pi+2}{4\pi r^2_s\nu}q^2_0\right)q_2+\\
&+\left(\frac{3}{2^{17/4}r_s^{5/2}\nu^{3/2}}q_2-\frac{1}{5\cdot2^{3/4}r_s^{3/2}\nu^{1/2}}\right)q_0^{5/2}+\\
&+\left(\frac{r^2_s\nu^2-1}{12\pi\sqrt{2}r^3_s\nu^2}q_2+\frac{\pi+2}{12\pi r^2_s\nu}\right)q_0^3-\frac{r^2_s\nu^2-1}{24\pi\sqrt{2}r^3_s\nu^2}q^4_0-\\
&-\frac{9}{7\cdot2^{17/4}r^{5/2}_s\nu^{3/2}}q^{7/2}_0-\frac{2(4\nu-1)}{q^2_2}\left(1-\frac{q_0}{2q_2}\right),
\end{split}
\end{equation}
where
\begin{equation*}
q_2=2\left(\frac{4\nu-1}{|I(q_0)|}\right)^{1/3}.
\end{equation*}
At $\nu\leq3$ and $1\lesssim r_s\lesssim2$,
\begin{equation}\label{q0}
q_0=\frac{9\times2^{1/4}\widetilde{r}_s^2-3\pi\widetilde{r}_s^{3/2}+\frac{15\pi}{2^{7/2}}\widetilde{r}_s^{1/2}-2^{1/4}}
{2^{5/4}\widetilde{r}_s^2+3\pi\widetilde{r}_s^{3/2}-2^{7/4}(\pi+2)\widetilde{r}_s+\frac{45\pi}{2^{7/2}}\widetilde{r}_s^{1/2}-2^{5/4}}.
\end{equation}
where $\widetilde{r}_s=\nu r_s$; i.e., $q_0\approx1$. At large $\nu$ values, according to Eq. \eqref{Ecorr},
\begin{equation}\label{Ecorr_big_nu_estimate}
E_\text{cor}\gtrsim-4\left(\frac{6}{\pi}\right)^{1/3}n_\text{2D}^{1/3}.
\end{equation}
Here, we take into account that, at $\nu\gg1$, the position of the minimum of the function $I(q)$ appreciably deviates from unity ($q_0\rightarrow2\sqrt{2}$).

The comparison of the dependence of the correlation energy on the number of valleys calculated by Eq. \eqref{Ecorr} with that obtained taking into account the
first and second order corrections with respect to the deviation of $q_1$ from $q_0$ is shown in Fig.~1 for the case of $\sigma=1$ and $n_\text{2D}=1/\pi$ ($r_s=\sqrt{\nu}$). It is noteworthy that the corrections to $E_\text{cor}$ only slightly affect the results (red asterisks almost coincide with black points). At large $\nu$ values, the correlation energy tends to estimate \eqref{Ecorr_big_nu_estimate}. The same figure shows the numerically calculated dependence of the energy of the ground state $E_\text{gs}$ on $\nu$.

\begin{figure}[t!]
\begin{center}
\includegraphics[width=0.47\textwidth]{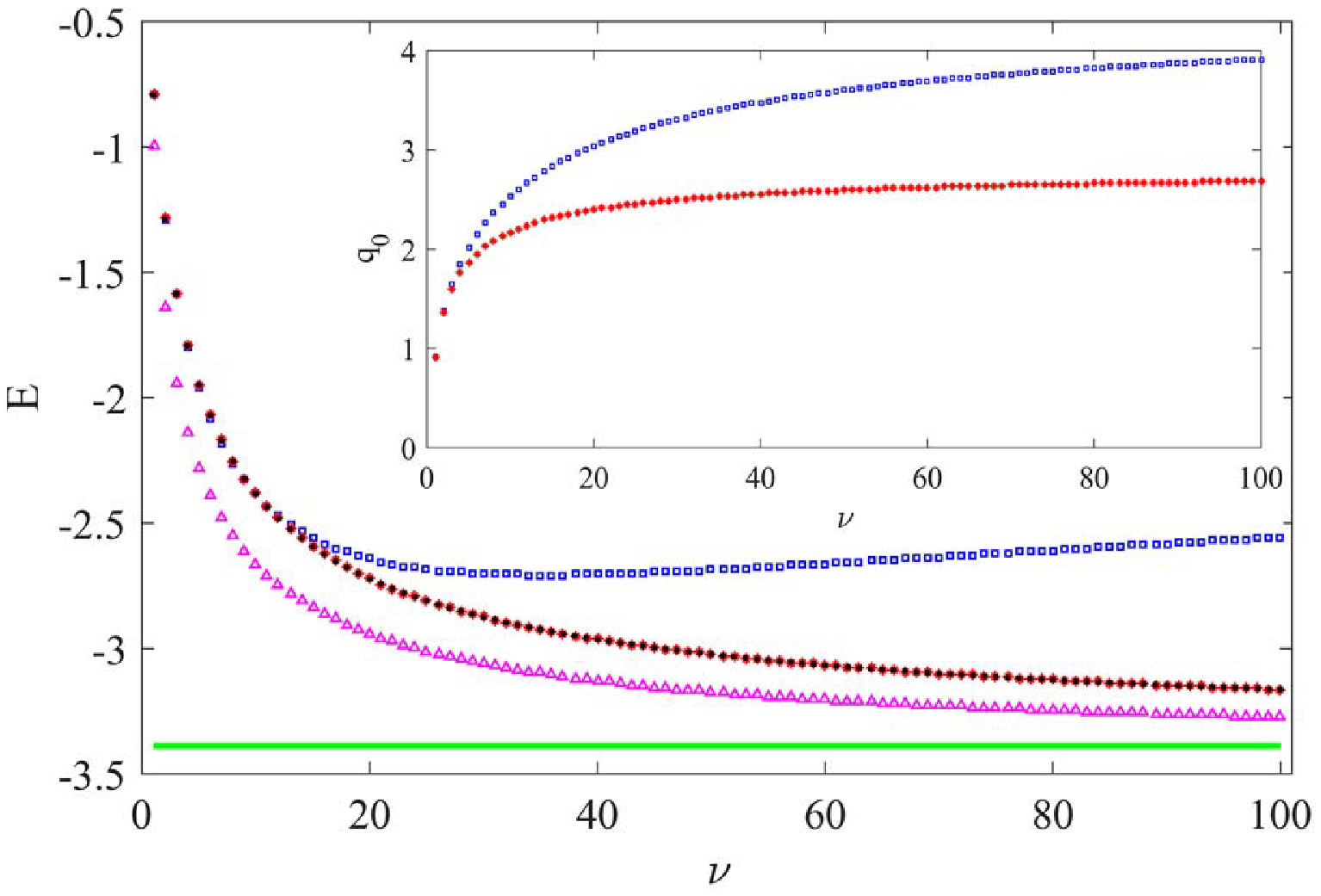}
\end{center}
\textbf{Fig. 1.} (Color online) Numerically calculated correlation energy and the energy of the ground state versus the number of valleys: (blue squares) $E_\text{cor}$ calculated by Eqs. \eqref{Ecorr} and \eqref{q0}, (red stars) $E_\text{cor}$ calculated by Eq. \eqref{Ecorr} with the numerical solution of the equation for $q_0$, and (black circles) $E_\text{cor}$ calculated taking into account corrections of the first and second orders with respect to the deviation of $q_1$ from $q_0$ ($q_0$ was determined numerically). The lower green horizontal straight line corresponds to estimate \eqref{Ecorr_big_nu_estimate} and purple triangles are $E_\text{gs}$ values calculated by Eqs. \eqref{E_gs}--\eqref{Eexch} and \eqref{Ecorr}. The inset shows the $q_0$ values (blue squares) calculated by Eq. \eqref{q0} and (red stars) calculated numerically. Note that $q_0\rightarrow2\sqrt{2}$ at $\nu\gg1$.
\end{figure}

\begin{center}
\textit{3.2. Calculation of the Correlation Energy in\\ the Limit of a Large Number of Valleys}
\end{center}

At $\nu\gg1$, when $n_\text{2D}$ satisfies inequalities \cite{Andryushin1976a, Andryushin1976b}
\begin{equation}\label{noneqvities}
1\ll q_F\ll n_\text{2D}^{1/4},
\end{equation}
the correlation energy is given by the expression
\begin{equation}\label{Ecorr_0}
E_\text{cor}=-\frac{1}{n_\text{2D}}\int\frac{d^2q}{(2\pi)^2}\int\limits_{-\infty}^\infty\frac{d\omega}{2\pi}\int\limits_0^1\frac{d\lambda}{\lambda}\mathcal{F}(\mathbf{q},\,\omega;\,\lambda).
\end{equation}
Here,
\begin{equation*}
\mathcal{F}(\mathbf{q},\,\omega;\,\lambda)=\frac{\lambda V(\mathbf{q})\Pi_0(\mathbf{q},\,i\omega)}{1-\lambda V(\mathbf{q})\Pi_0(\mathbf{q},\,i\omega)}-\lambda V(\mathbf{q})\Pi_0(\mathbf{q},\,i\omega),
\end{equation*}
where
\begin{equation}\label{Pi_big_q_omega}
\Pi_0(\mathbf{q},\,i\omega)=-2n_\text{2D}\sum\limits_{j=e,h}\frac{\varepsilon_j(\mathbf{q})}{\varepsilon_j^2(\mathbf{q})+\omega^2}
\end{equation}
is the polarization operator in the zeroth approximation with respect to the interaction constant at high momentum transfers ($q\gg q_F$) and frequencies ($\omega\gg E_F$). The dispersion relations are the same as in Eq.~\eqref{disp_lows}.

Relation \eqref{Ecorr_0} can be easily expressed in terms of the dimensionless units $q=\left(4\pi n_\text{2D}\lambda\right)^{1/3}\xi$ and $\omega=\left(4\pi n_\text{2D}\lambda\right)^{2/3}\zeta$. The correlation energy at an arbitrary ratio $\sigma$ of the electron and hole masses can be written as
\begin{equation}\label{Ecorr_big_nu}
E_\text{cor}=-A(\sigma)n^{1/3}_\text{2D},
\end{equation}
where
\begin{equation*}
A(\sigma)=\frac{3}{(4\pi)^{2/3}}\int\limits_0^\infty d\xi\int\limits_{-\infty}^\infty d\zeta\frac{\xi^3\left(\frac{\eta_e}{\xi^4+\eta_e^2\zeta^2}+\frac{\eta_h}{\xi^4+\eta_h^2\zeta^2}\right)^2}
{1+\xi\left[\frac{\eta_e}{\xi^4+\eta_e^2\zeta^2}+\frac{\eta_h}{\xi^4+\eta_h^2\zeta^2}\right]}.
\end{equation*}

Figure~2 shows the numerically calculated function $A(\sigma)$. It is convenient to fit it by the expression
\begin{equation*}
A(\sigma)\approx\frac{0.23}{\sigma^{2/3}}e^{-4\sigma}-0.098\sigma^3+0.378\sigma^2-0.442\sigma+4.932.
\end{equation*}

\begin{figure}[t!]
\begin{center}
\includegraphics[width=0.45\textwidth]{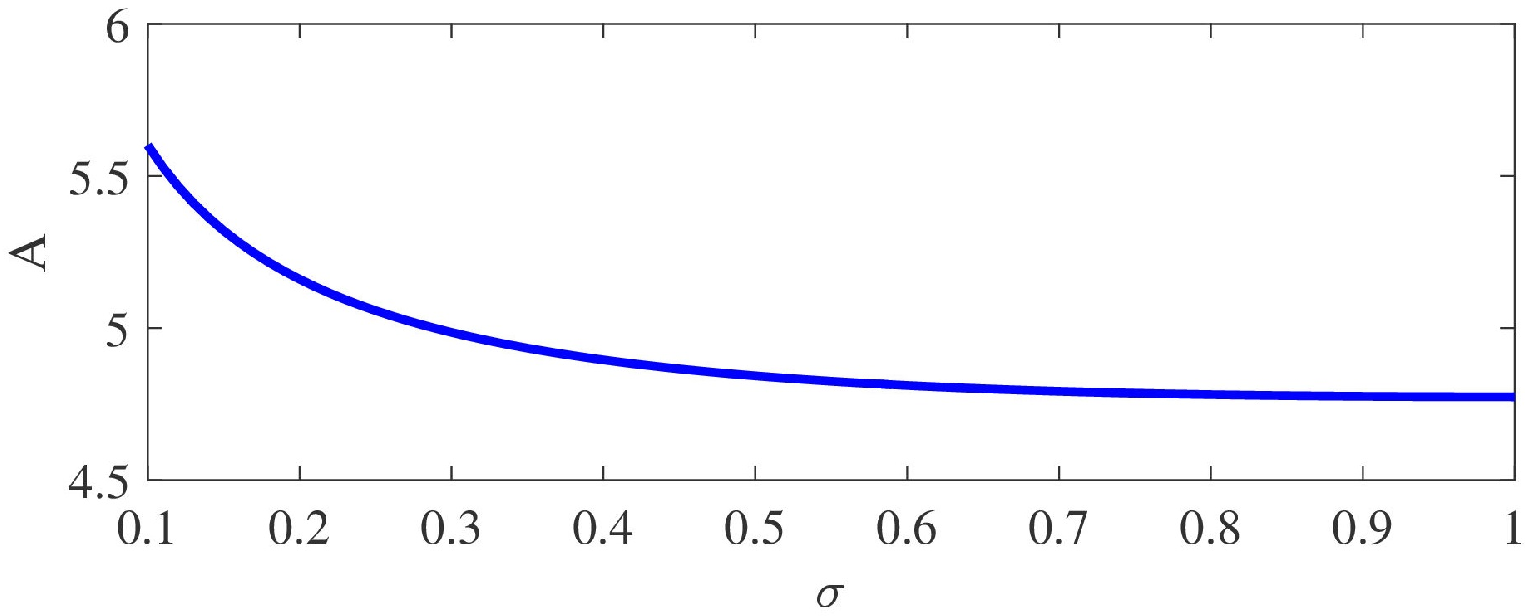}
\end{center}
\textbf{Fig. 2.} (Color online) Numerical calculation of $A(\sigma)$.
\end{figure}

For $\sigma=1$, we have
\begin{equation*}
A(1)=\frac{3\times2^{1/3}}{\pi^{1/6}}\Gamma\left(\frac{2}{3}\right)\Gamma\left(\frac{5}{6}\right)\approx4.774,
\end{equation*}
where $\Gamma(x)$ is the Euler gamma function.

This value is close to the factor in estimate \eqref{Ecorr_big_nu_estimate}. Note that estimate \eqref{Ecorr_big_nu_estimate} in the limit $\nu\rightarrow\infty$ remains a lower bound for \eqref{Ecorr_big_nu_estimate}. The constant $A(1)$ is smaller than the factor in estimate \eqref{Ecorr_big_nu_estimate} because the asymptotic expression \eqref{Pi_big_q_omega} incompletely includes the contribution of low momenta and frequencies.

\begin{center}
\textit{3.3. Calculation of the Ground State Energy\\
Using the Keldysh Potential}
\end{center}

The average kinetic energy is also given by Eq. \eqref{E_kin}. The exchange energy is expressed in the form ($\rho_0=r_0q_F$)
\begin{equation}\label{Eexch_K}
E^K_\text{exc}=-\frac{\sqrt{2}}{\pi r_s}\left[\frac{8}{3}-J(\rho_0)\right],
\end{equation}
where
\begin{equation*}
J(\rho_0)=\int\limits_0^1xdx\int\limits_0^1ydy\int\limits_0^{2\pi}\frac{\rho_0d\varphi}{1+\rho_0\sqrt{x^2+y^2-2xy\cos\varphi}}.
\end{equation*}
At characteristic densities $n_\text{2D}\sim10^{13}-10^{14}$ cm$^{-2}$, the dimensionless parameter is $\rho_0\simeq2-9$ at $\nu=2$. Therefore, it is interesting to calculate numerically the function $J(\rho_0)$ in the range $0<\rho_0<10$. In the limit of large $\rho_0$ values, it tends to 8/3 (see Fig.~3)

\begin{figure}[t!]
\begin{center}
\includegraphics[width=0.45\textwidth]{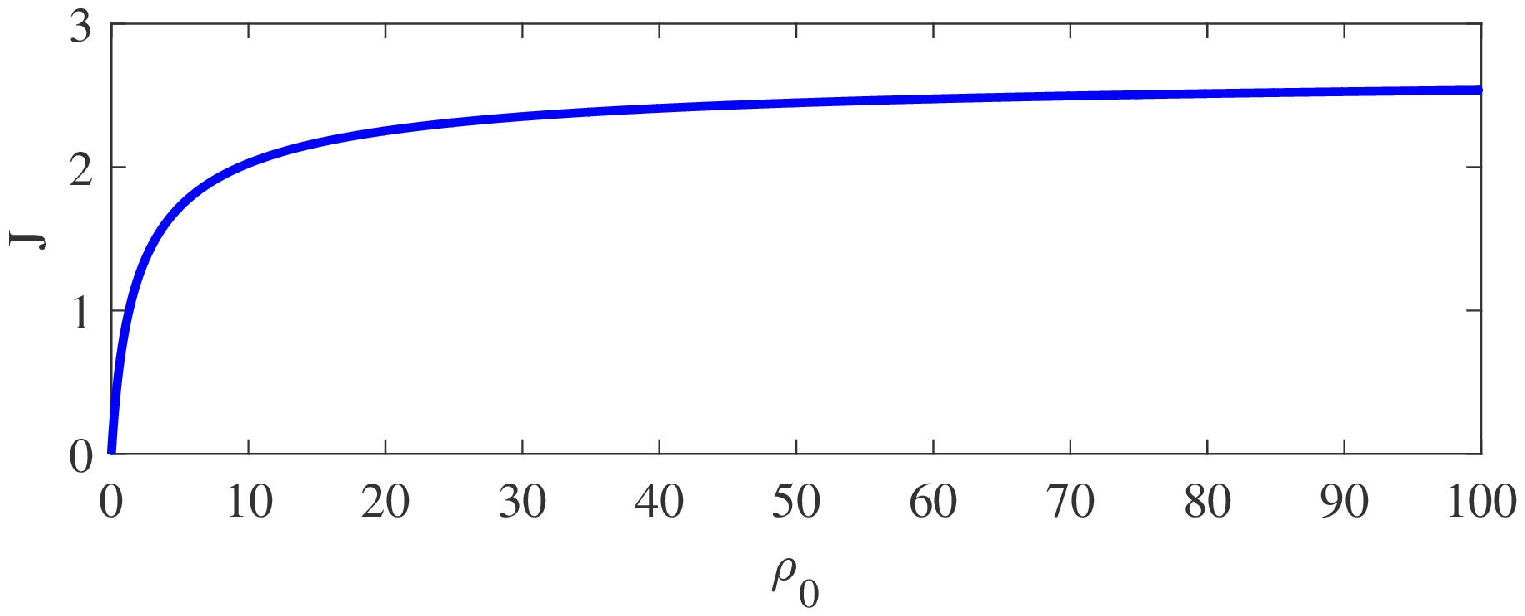}
\end{center}
\textbf{Fig. 3.} (Color online) Numerical calculation of $J(\rho_0)$.
\end{figure}

We determine the correlation energy by the method described in Subsection~3.1. Now, instead of the function $I(q)$, the integral in Eq. \eqref{corr_e} contains the function
\begin{equation}\label{I_q_K}
\widetilde{I}(q)=\begin{cases}
  -\frac{2\sqrt{2}}{\pi r_s(1+\rho_0q)}q+\frac{2^{1/4}}{r^{3/2}_s\nu^{1/2}\sqrt{1+\rho_0q}}q^{3/2}-&\\
  -\frac{\pi+2}{2\pi r^2_s\nu}q^2+\frac{3\sqrt{1+\rho_0q}}{2^{13/4}r^{5/2}_s\nu^{3/2}}q^{5/2}+&\\
  +\frac{r^2_s\nu^2-1}{6\pi\sqrt{2}r^3_s\nu^2(1+\rho_0q)}q^3, & \hspace{-1.25cm}q\ll1, \\
  -\frac{2(4\nu-1)}{q^3(1+\rho_0q)^2}, &  \hspace{-1.25cm}q\gg1.
\end{cases}
\end{equation}

In the intermediate range of wave vectors $q$, the function $\widetilde{I}(q)$ can also be fitted by a straight line segment. Integrating $\widetilde{I}(q)$ over $q$, we obtain an expression for the correlation energy $E^K_\text{cor}$. It is rather lengthy and is not presented here. However, for large $\nu$ values, when $\rho_0q_1\ll1$ and $\rho_0q_2\ll1$ ($q_1$ and $q_2$ are the matching points and $q_1\approx q_0$, where $q_0$ is the point of minimum of the function ), this expression can be expanded in powers of $\rho_0$ as
\begin{equation}\label{Ecorr_K}
E^K_\text{corr}=E_\text{cor}+\delta E^K_\text{cor},
\end{equation}
where $E_\text{cor}$ is given by Eq. \eqref{Ecorr}, whereas the correction in the approximation linear in $\rho_0$ reads
\begin{equation*}
\begin{split}
&\delta E^K_\text{corr}=\rho_0\left[-\frac{\sqrt{2}}{3\pi r_s}q^3_0+\frac{3}{7\cdot2^{7/4}r^{3/2}_s\nu^{1/2}}q^{7/2}_0+\right.\\
&\left.+\frac{2(4\nu-1)}{q_2}\left(3-\frac{q_0}{q_2}\right)+\left(\frac{\sqrt{2}}{\pi r_s}-\frac{\sqrt{q_0}}{2^{7/4}r^{3/2}_s\nu^{1/2}}\right)q^2_0q_2\right].
\end{split}
\end{equation*}

\begin{center}
4. EQUILIBRIUM DENSITY AND ENERGY\\
OF THE ELECTRON-HOLE LIQUID
\end{center}

The equilibrium density of the EHL $n_\text{EHL}$ is found by the minimization of the ground state energy. Substituting the correlation energy given by Eq. \eqref{Ecorr_big_nu} for a multivalley system into Eq. \eqref{E_gs} and taking the derivative with respect to $n_\text{2D}$, we obtain the following equation for $n_\text{EHL}$:
\begin{equation}\label{n_equilibrium_equation}
\left.\frac{\partial E_\text{gs}}{\partial n_\text{2D}}\right|_{n_\text{EHL}}=\frac{\pi}{\nu}-\frac{4\sqrt{2}}{3\sqrt{\pi\nu}}n^{-1/2}_\text{EHL}-\frac{1}{3}A(\sigma)n^{-2/3}_\text{EHL}=0.
\end{equation}
To solve Eq. \eqref{n_equilibrium_equation}, we note that the absolute value of the exchange energy given by Eq. \eqref{Eexch} at $\nu\gg1$ is less
than the absolute values of the kinetic and correlation energies. Therefore, we can first neglect the second term in Eq. \eqref{n_equilibrium_equation} and then find the exchange energy correction to the equilibrium density:
\begin{equation}\label{n_equilibrium}
n_\text{EHL}=\left(1+\frac{1}{1+\frac{1}{\sqrt{2}}\left(\frac{\pi}{3}\right)^{3/4}\nu^{1/4}A^{3/4}}\right)\left(\frac{\nu A}{3\pi}\right)^{3/2},
\end{equation}
\begin{equation}\label{E_equilibrium}
E_\text{EHL}=-\frac{2}{3}\left(\frac{\nu}{3\pi}\right)^{1/2}A^{3/2}-\frac{2^{7/2}\nu^{1/4}}{3^{7/4}\pi^{5/4}}A^{3/4}.
\end{equation}

In our opinion, multilayer multivalley systems are promising (the number of layers is $l$ and the distance between them is $c$). If $d=lc\lesssim a_x$, we can neglect the
second term in the denominator of Eq. \eqref{keldysh_int}. In this case, Eq. \eqref{coloumb_2d} can be used. The effective number of valleys $\nu_\text{eff}=l\nu$ of the system increases significantly.

The equilibrium density and energy of the EHL calculated using Eqs. \eqref{Ecorr} and \eqref{Ecorr_big_nu} for the correlation energy in the case of 10 separated TMD monolayers ($\nu_\text{eff}=20$) are shown in Fig.~4 as a function of the ratio of electron and hole masses.

\begin{figure}[t!]
\begin{center}
\includegraphics[width=0.48\textwidth]{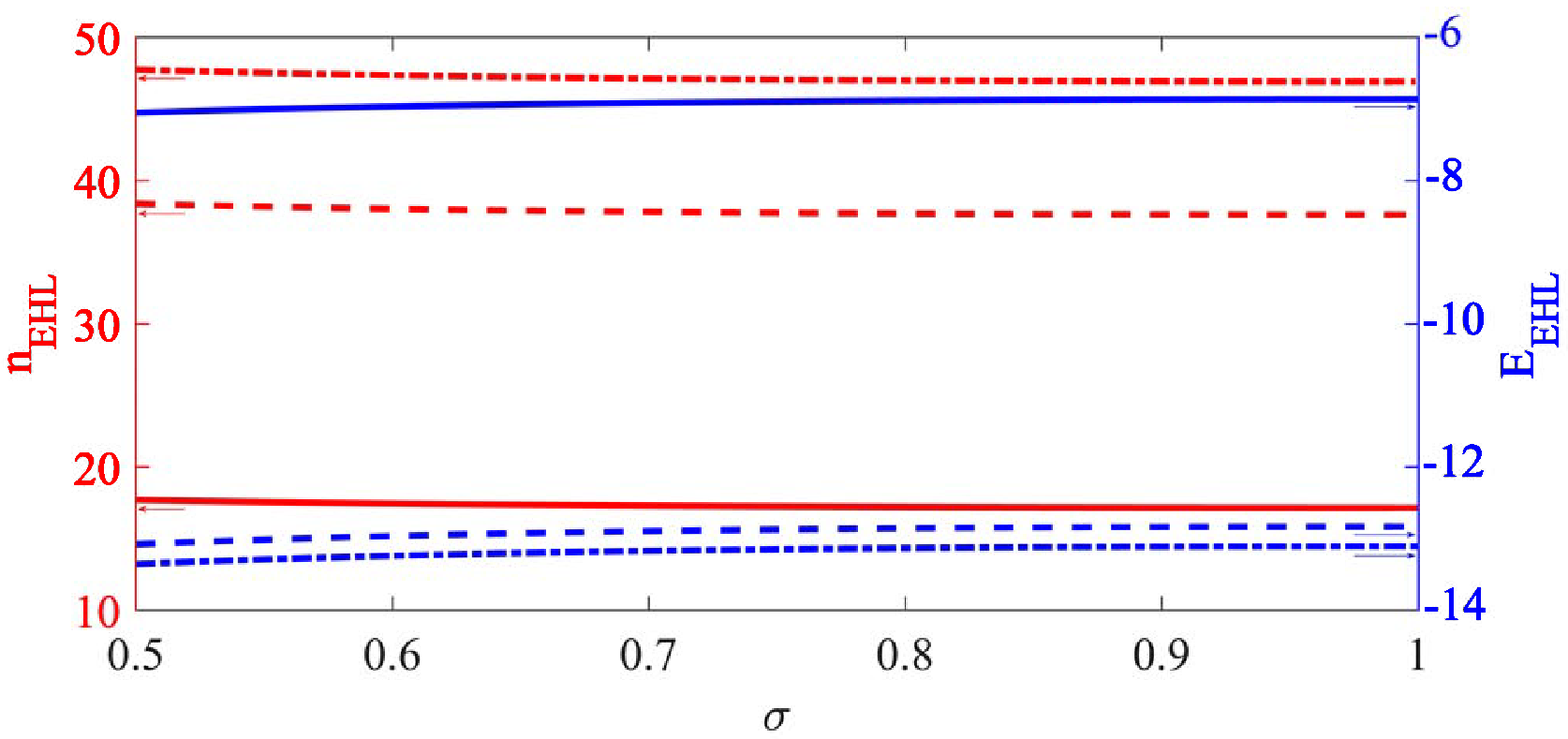}
\end{center}
\textbf{Fig. 4.} (Color online) (Red color) $n_\text{EHL}$ and (blue color) $E_\text{EHL}$ calculated for the 10-layer array using (solid lines) Eq. \eqref{Ecorr} and (dashed and dash-dotted lines) Eq. \eqref{Ecorr_big_nu} with (dashed lines) Eqs. \eqref{n_equilibrium} and \eqref{E_equilibrium} and (dash-dotted lines) the numerical solution of Eq.~\eqref{n_equilibrium_equation}.
\end{figure}

\begin{center}
5. COMPARISON OF THE CALCULATIONS\\
WITH EXPERIMENTAL DATA
\end{center}

Let us compare the results calculated by Eqs. \eqref{E_gs}--\eqref{Eexch} and \eqref{Ecorr} with the experimental data for $n_\text{EHL}$, $|E_\text{EHL}|$, and $T_c$ obtained for the monolayer MoS$_2$ film \cite{Yu2019}. With a good accuracy, we can assume that $m_e\approx m_h$~\cite{Eknapakul2014}. If $\nu=2$, we have $|E_\text{EHL}|=700$ meV, $n_\text{EHL}=10^{14}$ cm$^{-2}$ and $T_c\simeq800$~K. The corresponding experimental data are $|E_\text{EHL}|=480$ meV, $n_\text{EHL}=4\times10^{13}$ cm$^{-2}$, and $T_c\simeq500$ K \cite{Yu2019}. We can give two explanations of this discrepancy.

\textbf{The first explanation.} The number of valleys decreases. Stresses arising in a monolayer film can lift the degeneracy of valleys \cite{Sibeldin2016, Sibeldin2017}. In addition, the lifting of spin degeneracy of charge carriers is also equivalent to halving the number of valleys. This is possible due to the large spin-orbit splitting of the valence band, $\Delta_{vb}\approx148$~meV \cite{Rustagi2018}. For $\nu=1$, we obtain good agreement with the experiment: $|E_\text{EHL}|=450$ meV,
$n_\text{EHL}=3.3\times10^{13}$ cm$^{-2}$, and $T_c=520$~K.

\textbf{The second explanation.} It is necessary to use the Keldysh potential, which contains the fitting parameter $r_0$. At $r_0=0.7$ \AA\, and $\nu=2$, the calculated EHL energy $|E_\text{EHL}|=480$ meV is in the best agreement with its experimental value, but the density of the EHL is
overestimated: $n_\text{EHL}=5.4\times10^{13}$ cm$^{-2}$.

In the quantitative description of the positions of exciton lines in the photoluminescence spectrum of the MoS$_2$ monolayer, we find $r_0=41.47$ \AA\, \cite{Berkelbach2013}. The discrepancy in the values of $r_0$ is large because the calculations of excitons and EHL involve quite different ladder and loop diagrams, respectively.

We favor the first explanation.

\begin{center}
6. CONCLUSIONS
\end{center}

To summarize, we have obtained analytical and numerical results for the binding energy of the EHL and its equilibrium density in 2D systems with TMD monolayers at an arbitrary number of valleys.

We have also calculated the characteristics of the EHL using the Keldysh potential. It turned out that using only one parameter does not simultaneously match the binding energy of the EHL and its equilibrium density with experimental results. This suggests a quite limited range of applicability for the Keldysh potential in these calculations.

The difference between the theoretical and experimental results is due to the insufficient accuracy of the used parameters of the monolayer heterostructures, to the use of experimental $E_x$ and $a_x$ values as the units of measure, and to the possible inhomogeneity of the sample.

\small{
\begin{center}
FUNDING
\end{center}

P.\,V. Ratnikov acknowledges the support of the Foundation for the Advancement of Theoretical Physics and Mathematics BASIS (project no. 17-14-440-1, the general formulation of the problem) and of the Russian Science Foundation (project no. 16-12-10538-$\Pi$, the calculations of the correlation energy, Section~3).
}

%\newpage

\begin{flushright}
\textit{Translated by K. Kugel}
\end{flushright}
\end{document}